\documentstyle[aps,preprint]{revtex}
\tighten
\begin{document}
\draft
\hsize\textwidth\columnwidth\hsize\csname @twocolumnfalse\endcsname

\title{Universal Relation Connecting Fermi Surface to
Symmetry of the Gap Function in BCS-Like Superconductors}
\author{Gang Su$^{\ast}$ and Masuo Suzuki$^{\dag}$}
\address{ Department of Applied Physics, Faculty of Science,
 Science University of Tokyo\\
1-3, Kagurazaka, Shinjuku-ku, Tokyo 162, Japan}
\date{May 7, 1998}

\maketitle

\begin{abstract}
A universal relation connecting Fermi surface (FS) to the
symmetry of the
gap function in BCS-like superconductors is derived.
It is found that the shape of the FS can be deduced directly
from the symmetry of the superconducting gap function,
and is also influenced by the next nearest-neighbor overlapping.
The application of this relation to cuprate superconductors
is discussed. There is observed an interesting property
that Luttinger's theorem perfectly holds for the
tight-binding band while it is violated by inclusion of
the next nearest-neighbor overlapping integral.\\

\end{abstract}


A number of recent experiments\cite{photo,nmr,neutron,c-axis}
on underdoped bilayer cuprate superconductors reveal that
there appears a gap characterized by a large suppression of low
frequency (spin and electronic) spectral weight
near $(0,\pi)$ in the
normal state. This unusual phenomenon is commonly referred to
as the pseudogap (or spin gap,
see, e.g. Refs. \cite{randeria,maple}
for a review). Several scenarios, e.g. SU(2) gauge theory\cite{lee},
spinon pairing\cite{spinon}, pairing correlations\cite{randeria1},
superconducting (SC) fluctuations\cite{ioffe}
and so on, are thereby proposed to explain this unusual
phenomenon. In most of these theories, the Fermi
surface (FS) plays a great role. The interest in studying theories
of the FS is thus renewed.
The question how the FS is related
to the symmetry of the SC gap function, namely a basic issue
for addressing the cause of the pseudogap
phenomenon in our opinion, is still not quite
clear and calls for explorations.
Once we obtain the knowledge about the FS at zero temperature,
we can study the temperature evolution\cite{norman}
of the FS from this surface.

According to Luttinger\cite{luttinger}, the FS can be defined as the
locus in momentum space where the renormalized single-particle energy
is equal to the chemical potential at zero temperature. In other
words, if we know the zero-temperature chemical potential of the system,
we know the shape of the FS. For superconductors when we lower the
temperature down to zero, the system goes into the SC ground state.
Considering that superconductivity develops directly from a metal,
one may expect that the FS might have a close relation
to the symmetry of the SC gap function.
Since the symmetry of the SC gap function,
e.g. in cuprate superconductors,
can be clearly detected in experiments, it might offer us
an opportunity to
address the aforementioned question. Along this line,
we shall write down in the present paper the general BCS-like SC
ground-state wave function
in which the gap function and the chemical potential are
explicitly included. On account of it,
a universal relation between the Fermi
energy and the symmetry of the SC gap function can be obtained. When
comparing it with cuprate superconductors, one could in turn gain
insight into the structure of the SC ground state of cuprates.
Generally speaking, for a given system the FS is also given, and
by incorporating the given Hamiltonian one can
determine the SC gap function from this FS. Our logic here
is, however, inverse. In spite of the reasons above mentioned,
our motivation also comes from the following two aspects.
First,  there is, at present, no consensus on the
model Hamiltonian for cuprate superconductors,
and meanwhile the controversial conclusions have been usually drawn
in literature owing to the use of different Hamiltonians.
In order to get reliable consequences in this situation,
an alternative starting point might be
Hamiltonian-free. Yet, the structure of the SC ground state would
be the same, as detected in experiments, no matter what kind of
Hamiltonians are assumed. Even if one is able to write
down a Hamiltonian proper for cuprates, it is also not possible
to obtain the exact SC ground state
owing to complexity of many-body problems, like the celebrated BCS
theory\cite{bcs} in which the proposed wave function
is neither an exact ground state nor an eigenstate of the
electron-phonon interacting Hamiltonian. In fact, many people have
used  BCS-like schemes to construct
their models for cuprate superconductors.
Second, since the SC gap function in cuprates is already known,
while the shape of the FS is now actively under debate,
it appears possible to get some information of the FS
directly from the gap function. Furthermore,
one can use the obtained results to compare with experiments to
examine whether the SC ground states
of cuprates are  of BCS-like form or not.

{\it SC wave function.} --- It has been well established from the
experiments like flux quantization, Andreev reflection, Josephson
effect, etc. that electrons responsible for superconductivity in
cuprates are paired in singlet spin states (see, e.g.
Ref.\cite{batlogg}), although the mechanism leading to electron
pairing is still quite controversial.
Therefore, the SC state can be considered as the
condensation of these singlet electron pairs, implying that
the corresponding SC wave function would be some kind of coherent
superpositions of electron pairs. In addition, a number of
measurements showed that the pairing symmetry in cuprates, namely
the SC gap function over the FS, is anisotropic and
has the $d_{x^2 -y^2}$-wave character, demanding that any promising
SC wave function should reflect this property. Considering the
remarkable fact that the most successful pairing wave
function up to date is BCS-like\cite{bcs},
we can first assume that
the SC state of cuprate superconductors could also have a form
similar to the BCS state, and then we observe if it is really so.
The following discussion is based on this viewpoint.
Before presenting our result, let us first give some preliminary
definitions. The singlet
electron pairing operator $b_{{\bf k}}^{\dagger}$ can be generally
expressed in momentum space as
\begin{eqnarray}
b_{{\bf k}}^{\dagger} = c_{{\bf k}+{\bf Q}\uparrow}^{\dagger}
c_{-{\bf k}\downarrow}^{\dagger},
\label{b-op}
\end{eqnarray}
where $c_{{\bf k}\uparrow}^{\dagger}$ denotes the creation operator
of a spin-up electron with momentum ${\bf k}$, and ${\bf Q}$ is the
total momentum of an electron pair. The SC gap function is usually
defined as
\begin{eqnarray}
\Delta_{{\bf k}}  = g({\bf k}) \Delta_{0}, ~~~\Delta_{0} = \langle
b_{{\bf k}}^{\dagger} \rangle,
\label{gap}
\end{eqnarray}
where $g({\bf k})$ characterizes the pairing symmetry.
The amplitude of the gap function, $\Delta_{0}$, depends generally
on ${\bf k}$ and the doping parameter $\delta$, and
can be in principle calculated explicitly from a given
Hamiltonian. In most cases,
$\Delta_{0}$ is often assumed to be ${\bf k}$
independent, like the SU(2) gauge theory developed in Refs.\cite{lee}.
Equation (\ref{gap}) comprises the
following cases. (1) Cooper pairing ($s$-wave): $g_{s}({\bf k}) =1$ and
${\bf Q} = (0, \cdots)$; (2) $\eta$ pairing: $g_{\eta}({\bf k}) =1$ and
${\bf Q} = (\pi, \cdots)$; (3) Extended $s$-wave pairing:
$g_{e-s}({\bf k}) = \cos k_{x} + \cos k_{y}$ and ${\bf Q} = (0, \cdots)$;
(4) $d_{x^2 -y^2}$-wave pairing: $g_{d_{x^2-y^2}}({\bf k}) =
\cos k_{x} - \cos k_{y}$ and ${\bf Q} = (0, \cdots)$;
(5) $d_{xy}$-wave pairing: $g_{d_{xy}}({\bf k})
= \sin k_{x}  \sin k_{y}$ and ${\bf Q} = (0, \cdots)$; and so forth.
For the $d_{x^2-y^2} + i s$ ( $d_{x^2-y^2} + i d_{xy}$)-wave mixing state,
the SC gap function can be written as
$\Delta_{{\bf k}} = \Delta_{d_{x^2-y^2}}
({\bf k}) + i \varepsilon \Delta_{s(d_{xy})}({\bf k})$ with $\varepsilon$
the fraction of $s(d_{xy})$ component mixed with the $d_{x^2-y^2}$ state
(see e.g. Ref.\cite{harlingen}).
Hence, almost all currently interesting pairing scenarios are covered
by this definition. With these definitions in mind,
we now write down the general BCS-like SC wave function
as the following form:
\begin{eqnarray}
|\Psi_{G} \rangle = A_{0} \exp (\sum_{{\bf k}} \xi (\{{\bf k}\})
c_{{\bf k}+{\bf Q}\uparrow}^{\dagger}c_{-{\bf
k}\downarrow}^{\dagger}) |0\rangle,
\label{ansatz}
\end{eqnarray}
where the variable $\xi (\{{\bf k}\})$ will be specified below,
$|0\rangle$ is the vacuum satisfying $c_{{\bf k}\uparrow}
(c_{{\bf k}\downarrow})|0\rangle =0$, and $A_{0}$ is a
normalization factor given by
\begin{eqnarray}
 A_{0} = \frac{1}{\sqrt{\prod_{{\bf k}} ( 1 + |\xi (\{{\bf k}\})|^2)}}.
\label{factor}
\end{eqnarray}
This form is quite natural, because Eq.(\ref{ansatz}) is nothing but
a coherent state of the electron pairs, and breaks the U(1) symmetry.
Considering the hard-core
property of the pairing operator $b_{{\bf k}}^{\dagger}$, namely,
$(b_{{\bf k}}^{\dagger})^2 =0$, one may find that
$|\Psi_{G}\rangle$ recovers the BCS form.
We note that some forms similar to Eq. ({\ref{ansatz})
have been discussed in a few textbooks, but the gap symmetry and the
pairing momentum are not particularly emphasized.
However, the present form is more general, because it includes almost all
pairing mechanisms with various pairing symmetries\cite{note}.
It turns out that Eq. (\ref{ansatz}) can serve as a reasonable
variational wave function for the SC ground state.
We would like to point out that unlike the RVB wave
function\cite{anderson} where the doubly-occupied sites are projected
out owing to the assumption of the t-J model, we here do not necessarily
do so, like the conventional BCS theory,
as we consider the problem directly in momentum space and need
not invoke {\it a priori} assumption of a large on-site Coulomb
repulsion, which is also consistent with the definition of
off-diagonal long-range
order (ODLRO) in which the ODLRO on off-site pairing is
strictly ruled out\cite{yang}.
Besides, since the type-II superconductors (e.g. cuprate superconductors)
are anisotropic, there are
vortex cores existing in the range between the lower
and the upper critical fields, suggesting that the equation
(\ref{ansatz}) when applied to this case, might not work.
However, we suppose that
Eq.(\ref{ansatz}) would be universal in the absence of an applied
field. [Above the lower
but below the upper critical field, Eq. (\ref{ansatz}) might
have other forms.]
In general, $\xi (\{{\bf k}\})$ is a functional
of the single-particle bare
dispersion $\epsilon_{{\bf k}}$, the chemical potential $\mu$,
the gap function $\Delta_{{\bf k}}$ and the doping parameter $\delta$:
\begin{eqnarray}
 \xi (\{{\bf k}\})  = \xi (\epsilon_{{\bf k}},
 \mu, \Delta_{{\bf k}}, \delta),
\label{xi-1}
\end{eqnarray}
where the $\delta$-dependence of $\xi (\{{\bf k}\})$
stemms from $\Delta_{0}$. If the Hamiltonian of the system is given,
$\xi (\{{\bf k}\})$ can be in principle determined variationally.
When applying Eq.(\ref{ansatz}) to Eq.(\ref{gap}) we get the following
self-consistent equation
\begin{eqnarray}
\Delta_{{\bf k}} = \frac{g({\bf k}) \xi (\{{\bf k}\})}
{1 + |\xi (\{{\bf k}\})|^2}.
\label{gap-xi}
\end{eqnarray}
This equation shows that the SC gap function is
relevant to the chemical potential, i.e., the FS.
Obviously, if we know an explicit form of $\xi (\{{\bf k}\})$,
then we can gain some information on the FS.

{\it Fermi Surface.} --- For the aim above mentioned,
we may assume that $\xi (\{{\bf k}\})$ takes the following
simplest form, i.e., the BCS-like form:
\begin{eqnarray}
 \xi (\{{\bf k}\})  =  \frac{\Delta_{{\bf k}}}{\epsilon_{{\bf
 k}}-\mu + \sqrt{(\epsilon_{{\bf k}} - \mu)^2 + |\Delta_{{\bf k}}|^2}},
\label{xi-2}
\end{eqnarray}
where we have scaled all the relevant energies with the overlapping
integral $t$ which was taken to be unity. It should be noticed that
any BCS-like mean-field theory in the weak-coupling ($\sim t$) limit
can lead to the universal form of
$\xi (\{{\bf k}\})$ like Eq. (\ref{xi-2}). Incorporating
Eqs.(\ref{gap-xi}) and (\ref{xi-2}), it gives rise to
\begin{eqnarray}
\mu = \epsilon_{{\bf k}} - g({\bf k}) \sqrt{\frac{1}{4} -
|\Delta_{0}|^2}.
\label{mu}
\end{eqnarray}
For a mixing state like $d_{x^2-y^2} + i s$ or $d_{x^2-y^2} + i d_{xy}$,
a similar calculation yields
$\mu = \epsilon_{{\bf k}} - {\tilde g}({\bf k}) \sqrt{\frac{1}{4} -
|\Delta_{0}|^2}$ with ${\tilde g}({\bf k}) = \sqrt{g_{d_{x^2-y^2}}
({\bf k})^2 + \varepsilon^2 g_{s(d_{xy})}({\bf k})^2}$.
From Eq.(\ref{gap-xi}) we know $|\Delta_{0}| \leq 1/2$, suggesting that
the property for Eq. (\ref{mu}) to be real is guaranteed.
Equation (\ref{mu}) shows that the zero-temperature chemical potential
(i.e., the Fermi energy), thereby the FS,
is closely related to the symmetry of the gap function
$g({\bf k})$, except $|\Delta_{0}| = 1/2$ where the term containing the
gap symmetry vanishes. This is subtle.
 At a first glance, this seems to be impossible, because
the existence of the SC gap leads to that
the FS has no definition in the SC state.
However, one should not forget that superconductivity derives from a metal,
and those electrons which are close to the FS
of the metal are paired
and responsible for superconductivity, while the
chemical potential enters into
the formalism as a constant, we therefore have Eq.(\ref{mu}).
In other words, {\it the FS determined by Eq. (\ref{mu})
should be understood as that of the metal from which superconductivity
develops}.
One may observe that for the BCS theory, $g({\bf k}) =1$,
the second term of Eq. (\ref{mu}) is a constant, showing
that the shape of the FS is not affected by the SC gap, and
is mainly controlled by the single-particle dispersion, as it should be.
While for cuprate superconductors, the gap symmetry
may be the $d_{x^2-y^2}$-wave. If the SC
state can be written as the form of Eq. (\ref{ansatz}),
then the shape of the
FS may be obtained directly from Eq. (\ref{mu}).
Experimentally, the
single-particle dispersion for cuprates has the form
\begin{eqnarray}
\epsilon_{{\bf k}}
= -2 (\cos k_x + \cos k_y ) - 4 t' \cos k_x \cos k_y,
\label{epsilon}
\end{eqnarray}
where we have taken $t=1$ as an energy scale for simplicity. Typically,
$t'=-0.05$ for LSCO and $-0.25$ for YBCO\cite{fulde}.
When we insert these data
into Eq. (\ref{mu}), we find that the shape of the FS for LSCO is
qualitatively different from that of experiments, while the FS for YBCO
looks quite similar to the ARPES determined one in optimal doping.
(Here we have assumed the $d_{x^2-y^2}$ symmetry and
$|\Delta_{0}| = 0.07$ which comes from the experimental
data $t \approx 0.5eV$ and $\Delta_{0} \approx 35meV$ in both cases.)
This implies that the structure of the
SC ground state for LSCO
could be different from that for YBCO, while the latter
might be more similar to BCS-like one.

Now let us discuss which parameter affects primarily the shape of the FS.
Equation (\ref{mu}) contains three parameters, namely $t'$,
$\Delta_{0}$ and $g({\bf k})$. For different values of $t'$,
we find that the shapes of FS change
drastically, as depicted in Fig. 1(a),
where we have taken $t' = -1/4, -1/8, 0,
1/8, 1/4$ with $\Delta_{0}=0.07$ and $g({\bf k}) =
\cos k_x - \cos k_y$. It
can be seen that the shape of the FS is  gradually closed from
$t'\leq 0$  (open) to  $t'>0$.
In Fig. 1(b), we show contours for different values of
$\Delta_{0} (=0, 0.1,
0.2, 0.3, 0.4, 0.5)$ with $t'=-0.25$ and
$g({\bf k}) = \cos k_x - \cos k_y$.
One can find that $\Delta_{0}$ does not play an
important role in determining
the shape of the FS.
Since the doping dependence of $\Delta_{{\bf k}}$ comes only
from $\Delta_{0}$, the present result shows that the doping does not have a
significant effect on the shape of the FS
in BCS-like superconductors, in sharp
constrast to the experimental observations in cuprate superconductors where
the evolution of the FS with the doping is clearly observed. This may imply
that the BCS-like SC state could not be applied
to the {\it whole} doping regime
in cuprates, but this does not rule out the possibility
that the BCS-like SC state can be applied to the cuprates
in certain fixed doping regime.
We also present the shapes of the FS for different gap symmetries with
$t'=0$, $\Delta_{0}=0.07$ and $\varepsilon =0.3$, as shown in Fig. 2.
We find that the FS
for the s-wave and extended s-wave symmetries are the same (contour a).
The reason is quite simple. For the s-wave symmetry, the second term of Eq.
({\ref{mu}) is a constant, while for the extended s-wave symmetry that
second term can be merged into the first term,
leading to the same shape of the FS as that of the s-wave. The FS for
the $d_{x^2 -y^2}$-wave symmetry is open, and composed of two separated
cosine-like curves (contours b),
while for the $d_{xy}$-wave symmetry the FS looks like a slightly deformed
and closed square (contour c).
For a mixing state $d_{x^2-y^2} + i d_{xy}$, we see that the FS
consists of four arcs (contours d), similar to (but not the same as)
the FS for the $d_{x^2-y^2}$-wave symmetry with $t'=-0.25$.
This demonstrates that the FS for the d-wave
symmetry differs from that for the s-wave, while it makes us not easy
to identify which state, the $d_{x^2-y^2}$ with next nearest-neighbor
overlapping or the $d_{x^2-y^2} +i d_{xy}$ with only tight-binding band,
is more suitable for cuprates like YBCO or BSCCO. Nonetheless,
we could remark that the shape of the FS in BCS-like superconductors
can be deduced directly from the symmetry of the SC gap function,
and is also influenced by the magnitude of
the next nearest-neighbor overlapping integral $t'$.

{\it Luttinger's Theorem.} --- Almost forty years ago on the basis of
an adiabatically perturbative expansion, Luttinger was able to show
that the volume (or area in two dimensions) enclosed by the
FS for interacting electrons is the same as that for noninteracting
electrons\cite{luttinger},
the assertion now known as Luttinger's theorem. Later on,
people find that this theorem indeed holds in most cases, but
exceptions were also found particularly in strongly correlated electrons,
where it was shown that this theorem was violated (see e.g.
Refs.\cite{altshuler}). How about Luttinger's theorem in
BCS-like superconductors? In accordance with Eq. (\ref{mu}), we draw
the FS for the noninteracting  and interacting cases, as shown in
Fig. 3(a), where we have taken $t'=0$, $\Delta_{0}=0.07$ and
$g({\bf k}) = \cos k_x - \cos k_y$. (Note that in this case
the FS for the noninteracting system is a square.)
After carefully checking the areas
enclosed by two surfaces, we find that the two areas enclosed by the FS
are exactly the same. We also checked other values of $\Delta_{0}$ for
different gap symmetries but keep $t'=0$, and the same result was observed,
namely, Luttinger's theorem perfectly holds in this case.
When we increase the magnitude of the next nearest-neighbor overlapping $t'$,
we find that  Luttinger's theorem becomes violated, even for a slight
tuning-on of $t'$.  Shown in Fig. 3(b) is for $t'=-0.05$,
$\Delta_{0}=0.07$ and $g({\bf k}) = \cos k_x - \cos k_y$.
From this figure, one can see
that the areas enclosed by the two surfaces are not equal, i.e., the area
enclosed by the FS (i.e. four arcs) for the noninteracting system
is larger than that for the interacting system. The present result reveals
that the inclusion of next nearest-neighbor overlapping integral,
a property of strongly correlated electrons, will violate Luttinger's
theorem, no matter what kind of gap symmetries (except the s-wave) are used,
implying no adiabatic connection between interacting and
noninteracting systems in this case.
Therefore, the present result is compatible with the
statement in Refs. \cite{altshuler}, i.e., strong correlations between
electrons might violate Luttinger's theorem.

In summary, we have derived a universal relation connecting
the Fermi surface
to the gap symmetry in BCS-like superconductors. On the basis of it, we
showed that the shape of the FS in BCS-like superconductors can be
deduced directly from the symmetry of the SC gap function, and is
affected by the next nearest-neighbor overlapping integral.
When this universal relation applied
to cuprate superconductors, we found that the BCS-like SC state
with $d_{x^2-y^2}$-wave symmetry could not
be suited to LSCO, while it is probably applicable to YBCO and BSCCO.
Finally, we observed that  Luttinger's theorem
perfectly holds for the tight-binding band,
while the inclusion of next nearest-neighbor overlapping would
violate this theorem, which is consistent with some previous
investigations
in strongly correlated electrons. Although our derivation is quite
simple, the result in our opinion contains essential physical
consequences. We expect that our presentation
could offer some bases for understanding the pseudogap phenomenon
and the temperature evolution of the Fermi surface in cuprate
superconductors.

\acknowledgments

One of authors (GS) is grateful to the Department of Applied Physics,
Science University of Tokyo, for the warm hospitality, and to the Japan
Society for the Promotion of Science (JSPS) for support.
This work has also been supported by the CREST (Core Research for
Evolutional Science and Technology) of the Japan
Science and Technology Corporation (JST).



\begin{figure}
Fig.1 The evolution of the Fermi surfaces with: (a) next
nearest-neighbor
overlapping integral $t' (= -1/4, -1/8, 0, 1/8, 1/4)$ with
$|\Delta_{0}|=0.07$; (b) the amplitude of the SC gap function
$|\Delta_{0}| (= 0, 0.1, 0.2, 0.3, 0.4, 0.5)$ with $t'=-0.25$. In both cases,
$g({\bf k}) = \cos k_x - \cos k_y$.

Fig.2 The shapes of the Fermi surface for different symmetries of the SC
gap function. Contour a: s-wave and extended s-wave; b: $d_{x^2 -y^2}$-wave;
c: $d_{xy}$-wave; d: $d_{x^2-y^2} + i d_{xy}$-wave.
 Here $t'=0$, $|\Delta_{0}|=0.07$ and $\varepsilon =0.3$.

Fig.3 Illustration of Luttinger's theorem (see text).
(a) $t'=0$; (b) $t'=-0.05$. In both cases, $|\Delta_{0}|=0.07$ and
$g({\bf k}) = \cos k_x - \cos k_y$.

\end{figure}

\end{document}